\documentclass{ws-mpla}
\usepackage[super]{cite}
\usepackage{graphicx}
\begin{document}

\markboth{Rebeca Gonzalez Suarez	}
{Recent CMS results in top and Higgs physics}

\catchline{}{}{}{}{}

\title{RECENT CMS RESULTS IN TOP AND HIGGS PHYSICS \footnote{Updated version of the Wine \& Cheese: Joint Experimental-Theoretical Physics Seminar ``Recent CMS results in top and Higgs physics'' given at Fermilab, 20 November 2015.}
}

\author{\footnotesize REBECA GONZALEZ SUAREZ\footnote{
CERN, CH-1211 Geneva 23, Switzerland}}

\address{Department of Physics and Astronomy, University of Nebraska-Lincoln\\
Lincoln, NE 68588-0299,
United States of America\\
rebeca@cern.ch}

\maketitle

\pub{Received (Day Month Year)}{Revised (Day Month Year)}

\begin{abstract}

After the Higgs boson discovery in 2012, the investigation of its properties and compatibility with the standard model predictions is central to the physics program of the LHC experiments. Likewise, the study of the top quark is still relevant at the LHC, more than two decades after its discovery at the Tevatron. Top quarks and Higgs bosons are produced at the LHC on a large scale and share a deep connection based on the large mass of the top quark. Both particles provide an excellent laboratory in which to search for new physics: the measurement of their properties tests the foundations of the standard model; and they feature prominently in a variety of exotic signals. The coupling of the Higgs boson to the top quark, a fundamental standard model parameter, can only be measured directly in processes where the two particles are produced together. The production of a Higgs boson together with one or two top quarks is also sensitive to several exciting new physics effects. A brief overview of the current experimental status of top quark and Higgs boson physics is presented using results from the CMS Collaboration.

\keywords{Top quark; Higgs boson; CMS Collaboration; LHC.}
\end{abstract}

\ccode{PACS Nos.: 14.80.Bn, 14.65.H, 13.85.-t, 12.60.-i}

\section{Introduction}	

The standard model (SM) of particle physics contains what we know so far about elementary particles and how they interact with each other. It is one of the most significant achievements of contemporary physics, with a descriptive and predictive power experimentally proven over the years with striking accuracy. 

Particle physics is a relatively young and fast-moving field. In about 100 years the experiments have evolved from very simple setups like the cloud chamber to large particle accelerators and colliders that take decades to build. Large international collaborations are needed to build and operate the experiments, as well as teams of theorists to develop the computing-intensive simulations and complex predictions required to study the data -a jump from the very small teams that worked at the inception of particle physics. 

Although most of the particles known were swiftly observed in the 1960s and 1970s, the observation of the top quark and the Higgs boson were long-term enterprises, due partly to the high collision energies needed to produce them. The quark model was proposed in 1964, the same year that Peter Higgs submitted a paper on the subject of symmetry breaking~\cite{Higgs1964}. The quark family was completed 31 years later, in 1995, when the CDF and D\O~Collaborations at the Tevatron observed the top quark~\cite{top1,top2}, and it was the year 2012 when CERN reported the discovery of the Higgs boson at the LHC~\cite{higgs1, higgs2}.

The Higgs boson is the first fundamental scalar particle observed, and a candidate to confirm the Higgs mechanism of mass generation via spontaneous electroweak (EWK) symmetry breaking. Since the Higgs boson discovery, the LHC experiments have found no other new particles or significant deviations from the SM predictions~\footnote{Small deviations, such as the one found in the lepton universality tests by LHCb~\cite{lhcb}, are not yet significant at the time of writing.}. Martin Lewis Perl, who won the Nobel Prize for the discovery of the $\tau$ lepton in 1995, said in 2014 that {\it`The easy elementary particle questions were answered thirty and forty years ago'}~\cite{Perl}. The LHC now has to confront the hard questions, such as the description of gravity; the origin of the neutrino masses; the matter-anti-matter imbalance of the Universe; and the nature of dark matter and dark energy. For this, physics beyond the SM (BSM) are needed. 

Finding new phenomena is therefore a central goal of the LHC experiments, and it can be pursued in two ways. The first is via direct searches, targeting a particular model or signature; the second concerns precision measurements of properties, where BSM effects can be discovered as deviations with respect to the SM predictions. The investigation of processes with top quarks and Higgs bosons, which are among the most interesting particles to study at the LHC today, appertains both categories.

In the following I will present selected results from the CMS Collaboration~\footnote{The ATLAS collaboration has compatible results.} that illustrate the current experimental landscape of top quark and Higgs boson physics. Both particles have remarkable and unique properties, and are interesting on their own; when considered together they are even more fascinating. I will briefly give an overview of the current status of Higgs boson physics first, followed by top quark physics, to then focus on the interplay of top and Higgs together. 

The LHC experiments are finishing the analysis of the Run~1 dataset, of about 5$\rm{fb^{-1}}$ at 7 TeV and up to 20$\rm{fb^{-1}}$ at 8 TeV, delivered between 2010 and 2012. The LHC Run~1 provided not only the first observation of the Higgs boson, but also the first characterization of its properties. 

The LHC entered a new energy regime in 2015, when the Run~2 started at 13~TeV. It will continue running until the end of 2018 and the full potential of the Run~2, expected to deliver about 100$\rm{fb^{-1}}$ by the end of 2018, is still ahead, but the LHC delivered already more than 40$\rm{fb^{-1}}$ of integrated luminosity in proton-proton collisions in 2016. The increase of center of mass energy together with the very large expected luminosity would substantially increase the potential of the LHC for finding new phenomena. 

\section{Higgs Boson Physics}

The start of the Run~1 of the LHC was also the kickoff of a thorough search for the Higgs boson by the two large experiments of the LHC. During the first two years the full mass range from the LEP limit of 114.4~GeV, up to well above 600~GeV, was explored. Since the main Higgs boson production mode at the LHC is gluon-gluon fusion independently of the Higgs boson mass considered, the main channels explored for observation depended directly on the Higgs boson decay. The Higgs boson was observed in three channels~\cite{hrun1}: H$\rightarrow$ZZ$\rightarrow$4l, that has a clear four-lepton invariant mass peak over a small continuum background and it is considered as `{\it the golden channel}' at the LHC; H$\rightarrow \gamma\gamma$, that also has a mass peak on a smooth background, that profits from high-resolution calorimetry; and H$\rightarrow$WW$\rightarrow 2l2\nu$, where the mass peak reconstruction is not possible but that has a clear signature and sizeable branching ratio at 125~GeV -the second after H$\rightarrow b\bar{b}$. Several other Higgs decay channels have been carefully explored, including the H$\rightarrow b\bar{b}$ channel, that suffers from high backgrounds and challenges presented by b-jet reconstruction, and the H$\rightarrow \tau\tau$ channel that requires a combination of multiple final state signatures. Sub-leading production modes, such as vector boson fusion, or associated production with a vector-boson, have also been studied.

\begin{figure}[h]
\centerline{\includegraphics[width=4.0in]{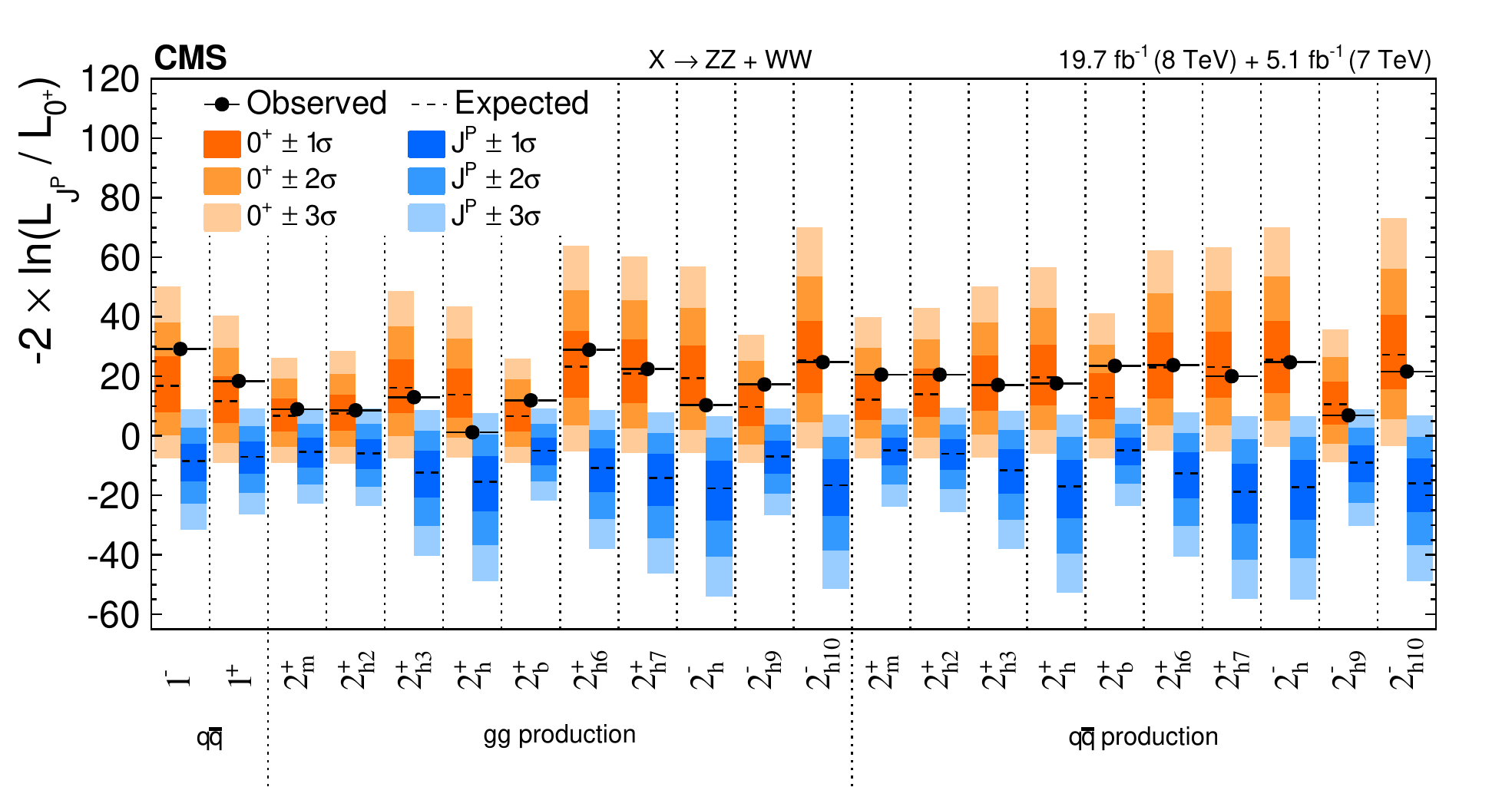}}
\vspace*{8pt}
\caption{Distributions of the test statistic $q=2ln(\mathcal{L}_{JP}/\mathcal{L}_{0+})$ for the spin-one and spin-two models tested against the SM Higgs boson hypothesis in the combined H$\rightarrow WW$ and H$\rightarrow ZZ$ analyses. The expected median and the 68.3\%, 95.4\%, and 99.7\% CL regions for the SM Higgs boson (orange, the left for each model) and for the alternative JP hypotheses (blue, right) are shown. The observed q values are indicated by the black dots~\cite{spin}.\protect\label{spinp}}
\end{figure}

The first studies of the Higgs boson properties were performed during Run~1. Decays of H$\rightarrow$VV (V=Z,W,$\gamma$) were used to explore the spin and tensor structure of the Higgs boson~\cite{spin}. The analysis used kinematic distributions of the decay products determined by the tensor structure of the HVV interactions. The number of kinematic variables available defined the strategy followed in each channel, and depends on the Higgs decay. The H$\rightarrow WW$ decay has limited sensitivity due to the undetected neutrinos in the dilepton final state. However for H$\rightarrow ZZ\rightarrow 4l$, the complete final state can be reconstructed and the full kinematic information is accessible and condensed in a set of discriminants. Different BSM spin and CP hypotheses were tested in the H$\rightarrow WW$ and H$\rightarrow ZZ$ processes, such as several exotic spin-1 or spin-2 scenarios and BSM $0^-$ and $0^+_h$. As it is shown in Fig.~\ref{spinp}, all spin-1 hypotheses have been excluded at more than 99.999\% confidence level (CL). The spin-2 boson $2^+_m$ was excluded at a 99.87\% CL, and other spin-2 hypotheses were excluded at 99\%~CL or higher. Therefore a spin 0 particle is favored by the data. 

The study of the tensor structure of the HVV interactions was performed under the spin-0 assumption, setting limits on anomalous couplings.  A parameterization of the decay amplitude of a spin-0 boson to a pair of V bosons was chosen, with three coefficients related to cross sections fractions. The analysis performed likelihood scans of sensitive kinematic variables for the effective fractions. A full set of scans in ZZ and WW decays is available, all consistent with the SM. The allowed confidence level intervals for anomalous coupling parameters assuming real coupling ratios ($\pi$ or 0) are all consistent with the SM as well. Assuming custodial symmetry, the pure $0_h^+$ hypothesis is excluded at 99.93\% CL and the pure $0^{-}$ (pseudoscalar) is excluded at 99.99\% CL. The analysis was repeated recently with Run~2 data in H$\rightarrow$ZZ$\rightarrow 4l$ decays~\cite{spinrun2}, and a preliminary result that includes additional Higgs production modes is available with a similar sensitivity.

\begin{figure}[h]
\centerline{\includegraphics[width=2.0in]{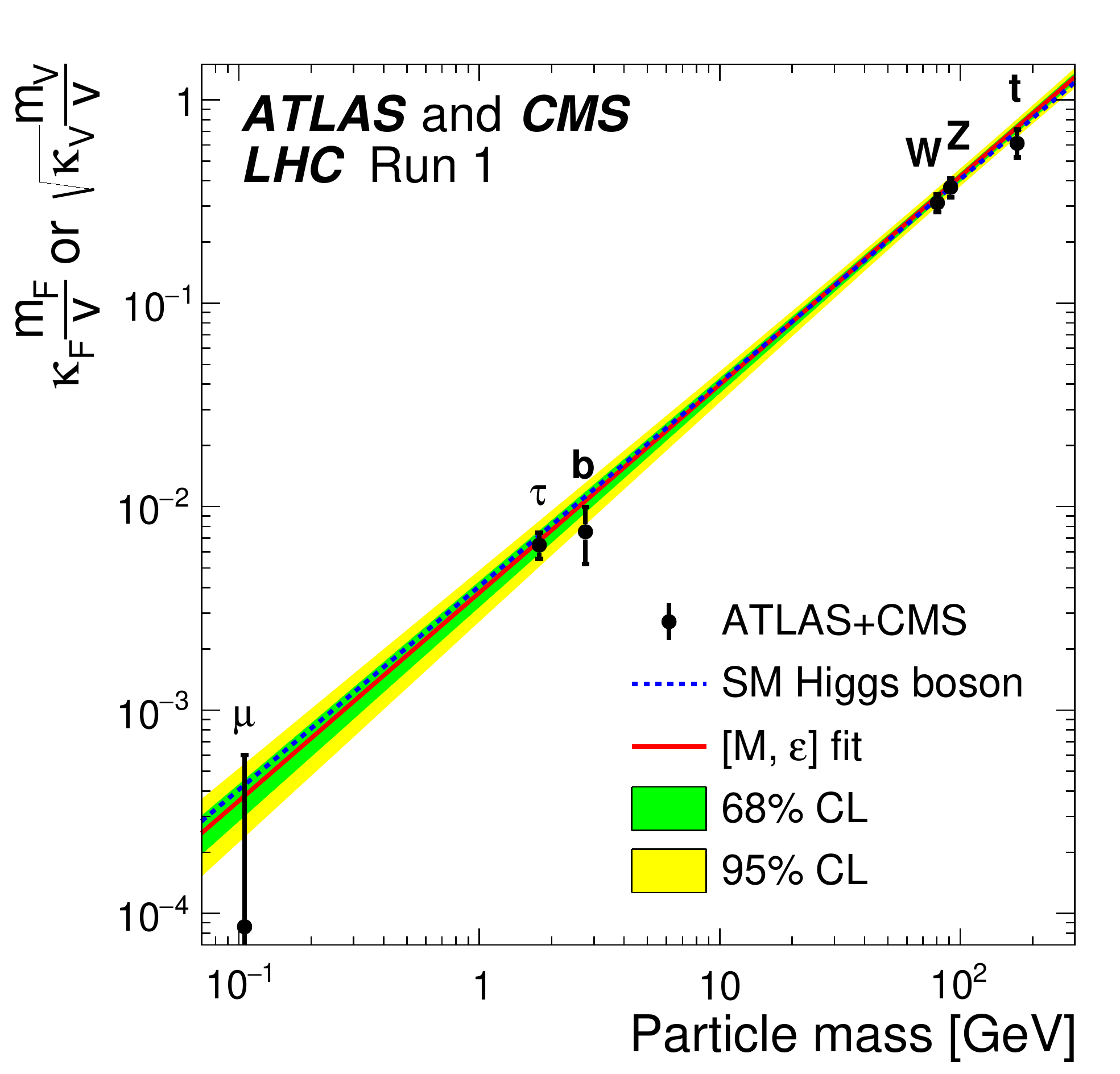}\includegraphics[width=2.0in]{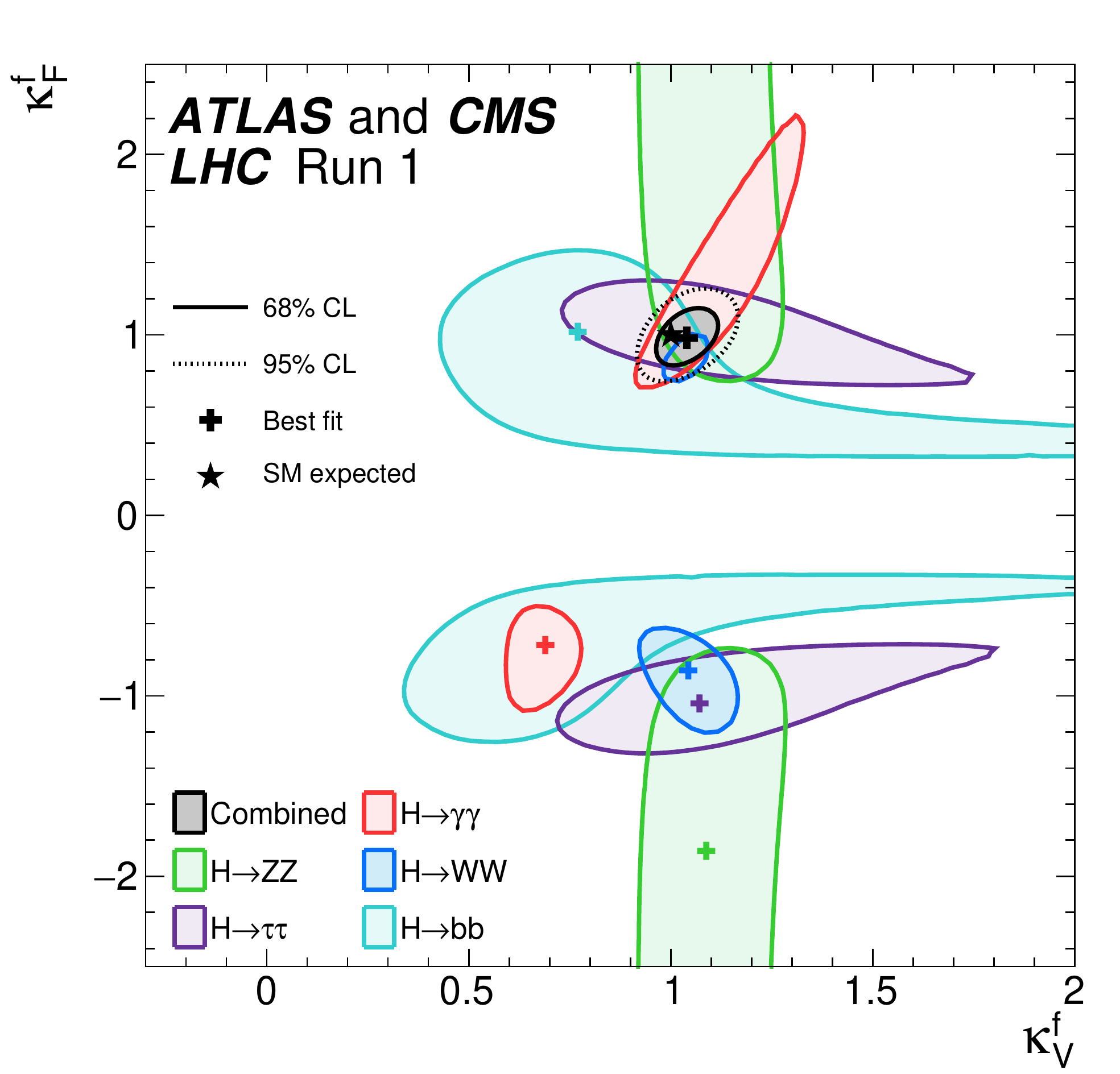}}
\vspace*{8pt}
\caption{Left: Best fit values of the Higgs boson couplings to the different particles as a function of particle mass for the combination of ATLAS and CMS data. Right: Negative log-likelihood contours at 68\% and 95\% CL in the ($\kappa_f$-$\kappa_V$, Higgs boson couplings to fermions and vector bosons) plane for the combination of ATLAS and CMS and for the individual decay channels, as well as for their combination~\cite{lhccoup}.\protect\label{hcop}}
\end{figure}

The CMS experiment performed a combined measurement of the mass of the Higgs boson with information from all the channels studied during Run~1~\cite{cmsmass}. In the high-resolution $\gamma\gamma$ and ZZ channels, the mass of the Higgs boson was determined to be: $\mathrm{m_H}=125.02\pm0.26\rm{(stat)}\pm0.14\rm{(syst)}$~GeV. The combined signal strength, defined as the measured cross section relative to the SM expectation at the measured mass, was found to be: $1.00\pm0.09\rm{(stat)}\pm0.08\rm{(theo)}\pm0.07\rm{(syst)}$. Compatibility of the Higgs boson couplings with the SM predictions was also explored~\cite{lhccoup} and, within uncertainties, all couplings are consistent with a SM Higgs boson, as can be seen in Fig.~\ref{hcop}.

\begin{figure}[h]
\centerline{\includegraphics[width=4.0in]{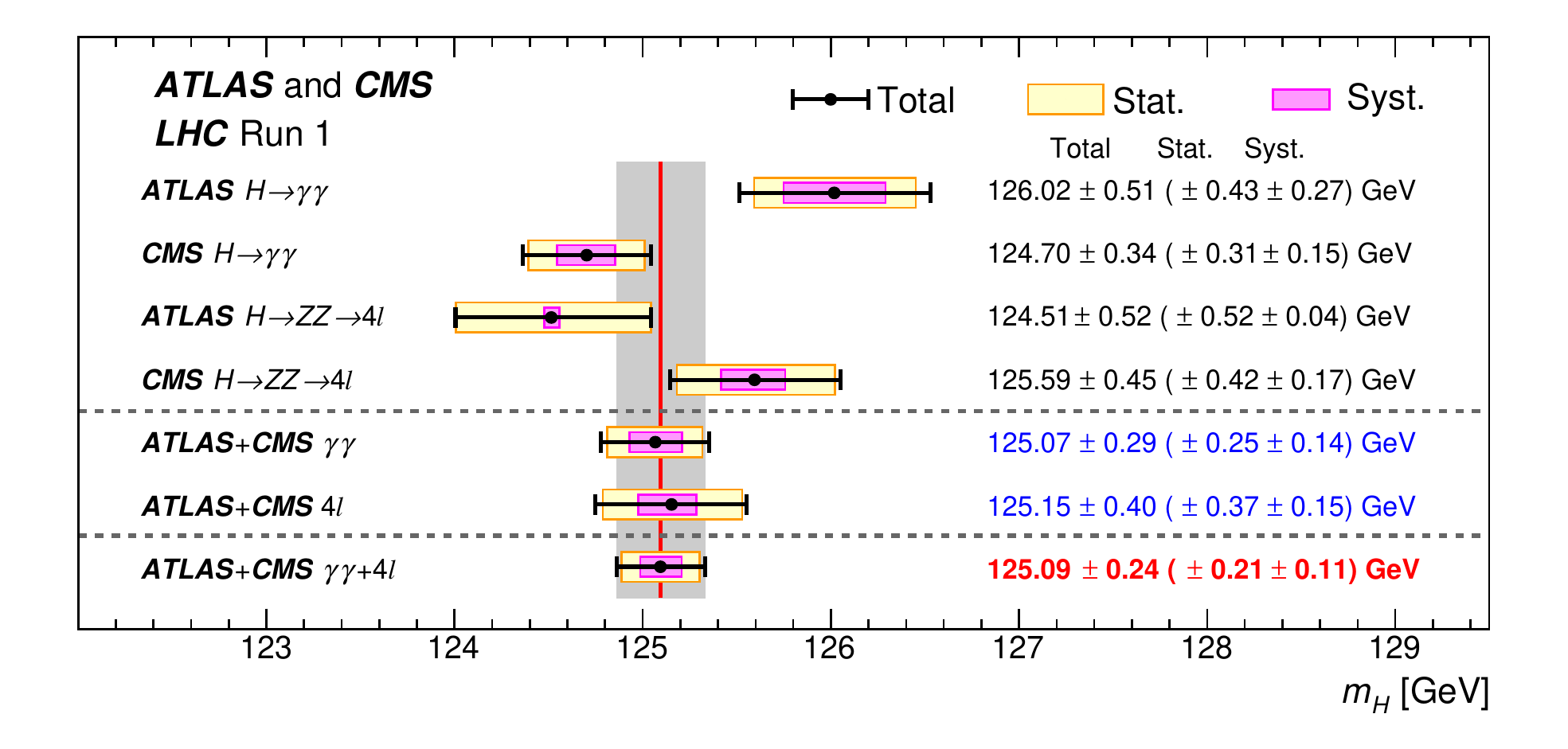}}
\vspace*{8pt}
\caption{Summary of Higgs boson mass measurements from the individual analyses of ATLAS and CMS and from the combined analysis. The systematic (narrower, magenta-shaded bands), statistical (wider, yellow-shaded bands), and total (black error bars) uncertainties are indicated. The (red) vertical line and corresponding (gray) shaded column indicate the central value and the total uncertainty of the combined measurement, respectively~\cite{lhcmass}.\protect\label{hmassp}}
\end{figure}

The LHC combination of the Run~1 measurements~\cite{lhcmass}, using also inputs from ATLAS, leads to improved precision for $\mathrm{m_H}$, on the order of 0.2\%. The result, $\mathrm{m_H}=125.09\pm0.21\rm{(stat.)}\pm0.11\rm{(syst.})$~GeV, and the different inputs are presented in Fig.~\ref{hmassp}.

The CMS collaboration also set limits on the Higgs boson lifetime and width from its decay to four charged leptons with Run~1 data~\cite{hwidth}. The measurement of the Higgs boson lifetime is derived from its flight distance in the CMS detector with an upper bound of $\tau_H<1.9\times10^{-13}$~s at the 95\% CL, corresponding to a lower bound on the width of $\Gamma_H >3.5\times10^{-9}$~MeV.  

With Run~2 data, several Higgs decay modes have already been explored and a number of preliminary results are available. Early measurements of Higgs boson properties are consistent with the SM expectations. For example, in the H$\rightarrow$ZZ$\rightarrow 4l$ channel~\cite{HIG-16-041}, the mass is measured to be $\rm{m_H}= 125.26\pm0.21$~GeV and the width, constrained using on-shell production, is found to be $<$1.10~GeV, at 95\% CL.

Therefore the new particle found at the LHC is a weak neutral resonance, with quantum numbers $\mathrm{J^{CP}}$=$0^{++}$, coupling strengths compatible with the SM Higgs boson, and a mass of 125.09~GeV, well within the allowed SM Higgs mass in the EWK fits. Some properties remain out of reach, for example the universality of the Higgs boson coupling to leptons, not accessible yet due to the low branching fraction of H$\rightarrow \mu\mu$ decays. The new particle discovered in 2012 looks then very much like a SM Higgs boson, but only more data will tell if there is something exotic in its nature. 

\section{Top Quark Physics}

With a mass of about 172~GeV, the top quark is the heaviest of all elementary particles. Due to its large mass, the top quark is very short-lived. In fact it is the only quark that decays before having time to form hadrons. This means that some of its properties, such as the spin and polarization, pass directly to its decay products. Therefore, by studying the top quark decay products it is possible to access some of its properties in a way that no other quark allows.

At the LHC, top quarks are produced mainly in $t\bar{t}$ pairs via strong interaction, mostly from gluons. There is an alternative mode that occurs at a lower rate, via EWK interaction: single top quark production. The production cross section of the different top quark processes at the Tevatron and the LHC is presented in Table~\ref{ta2}. Top quarks decay almost exclusively as t$\rightarrow$Wb, and then the W boson decays subsequently as $l\nu$ (32\%), or qq (67\%). Therefore, top quark final states are characterized by jets coming from b-quark decays and either one isolated lepton (e, $\mu$, $\tau$) and a neutrino, or a pair of light quarks. Final states with multiple top quarks therefore have very distinct experimental signals that use the full potential of the CMS experiment and all of its sub-detectors.

\begin{table}[h]
\tbl{Production cross section for different top processes at the Tevatron and the LHC.}
{\begin{tabular}{@{}lcccc@{}} \toprule
$\sigma$[pb] & $t\bar{t}$ & t-channel & tW & s-channel \\
\colrule
Tevatron 1.96~TeV &  7.08 & 2.08 & 0.22 & 1.046 \\
LHC 7~TeV &  177.31 & 63.89 & 15.74 & 4.29 \\
LHC 8~TeV & 252.89 &  84.69 & 22.2 & 5.24 \\
LHC 13~TeV &  831.76 & 216.99 & 71.2 & 10.32 \\ \botrule
\end{tabular}\label{ta2} }
\end{table}

The top quark legacy from the Run~1 of the LHC is vast, with more than 60 publications at the time of writing in CMS alone. It covers precise measurements of production cross sections, at the level of 3.5\%~\cite{xsec}; extensive differential~\cite{diff} studies, including boosted signatures; single top quark measurements, including the first observation of the tW production~\cite{tW} and properties in t-channel; first studies of very rare top quark processes, including four top quark production and the first observation of $t\bar{t}V$ (V=W,Z)~\cite{ttV}; and high precision measurements of top quark properties. Among the latter, the flagship is the top quark mass, $\mathrm{m_t}$. 

The most precise value of $\mathrm{m_t}$ to date, $\mathrm{m_t}=172.44\pm0.13\rm{(stat)}\pm0.47\rm{(syst)}$~GeV~\cite{topmass} was achieved by the CMS experiment by combining the main measurements performed during Run~1, with an uncertainty of 0.48~GeV (0.3\%). The top quark mass that is measured in classic analyses relates directly with the value of the top quark mass used in the simulation and has to be translated to theoretical fixed-order calculations of $\mathrm{m_t}$. At this level of precision, the discussion on how to perform this translation and the uncertainty associated with it becomes very important and requires large input from theorists. 

\begin{figure}[h]
\centerline{\includegraphics[width=4.0in]{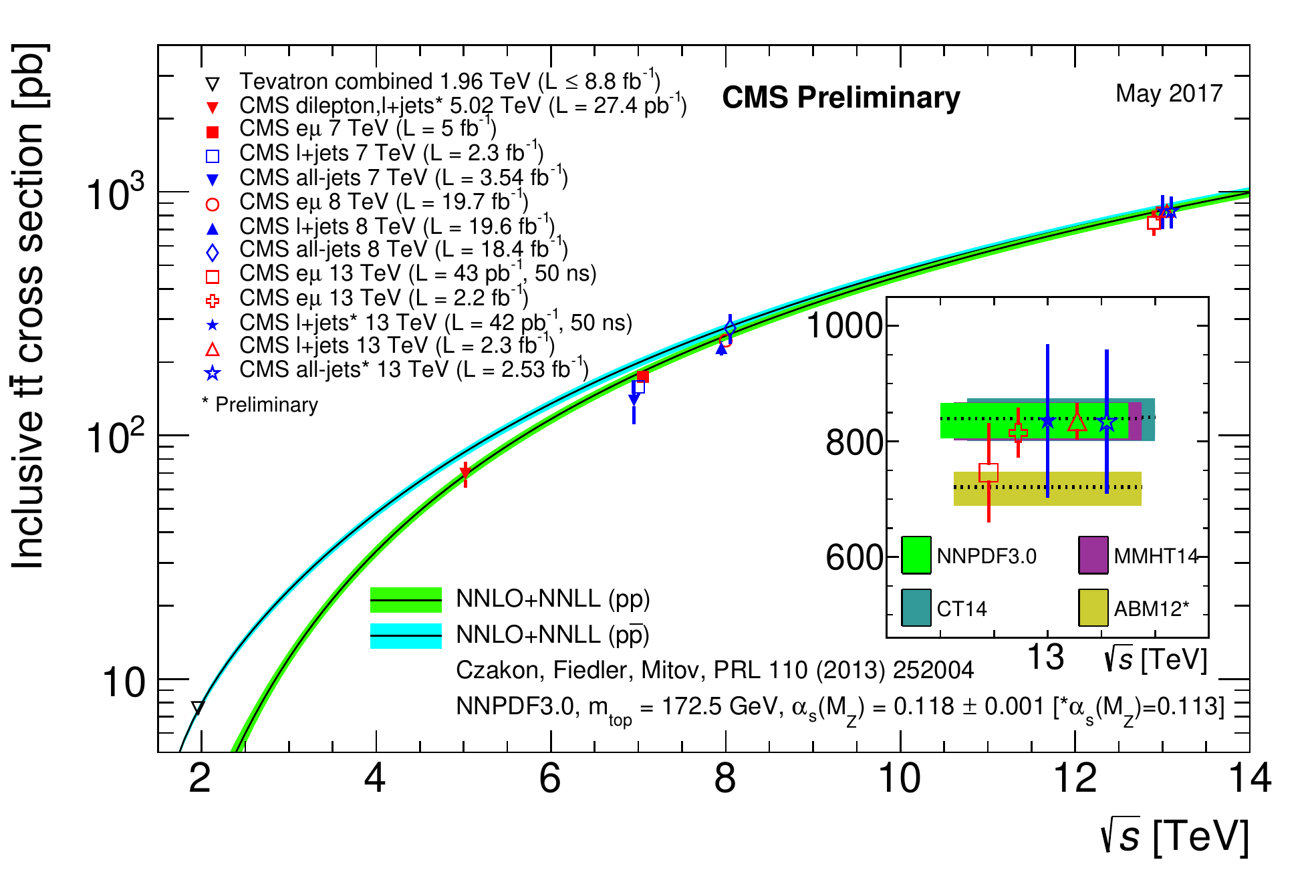}}
\vspace*{8pt}
\caption{Summary of inclusive $t\bar{t}$ production cross sections at different center of mass energies by the CMS Collaboration, including 13~TeV compared with NNLO+NNLL theory predictions.\protect\label{xsecp}}
\end{figure}

Already with the very early data of Run~2 the inclusive top pair production cross section was measured by CMS. In the dilepton final state (where both top quarks decay to leptonically decaying W bosons), with only 42$\mathrm{pb^{-1}}$ and a simple selection, the $t\bar{t}$ cross section was measured with less than 5\% of uncertainty~\cite{xsecEA}. Many other cross section measurements, using increasingly larger sets of dat, followed. At the moment of writing, there are measurements for the $t\bar{t}$ cross section using dilepton, lepton+jets, and all jets final states at 13~TeV, with uncertainties below 4\%. Figure~\ref{xsecp} presents a summary of some of the inclusive cross section measurements for $t\bar{t}$ production by the CMS Collaboration.

Differential measurements have also been performed with Run~2 data, first in the lepton+jets final state~\cite{diffR2}, which has the largest rate, followed by  dilepton and all jets final states. Distributions of global event variables (such as missing transverse momentum, or jet multiplicity) as well as distributions related to the $t\bar{t}$ system like the $p_T$ of the top quark, corrected for detector events to particle and parton level, are already available. Single top quark production, once a rare process that was observed for the first time in 2009, could already be observed also with the first 42$\mathrm{pb^{-1}}$ of collisions during Run~2~\cite{tchEA}. 

\begin{figure}[h]
\centerline{\includegraphics[width=4.0in]{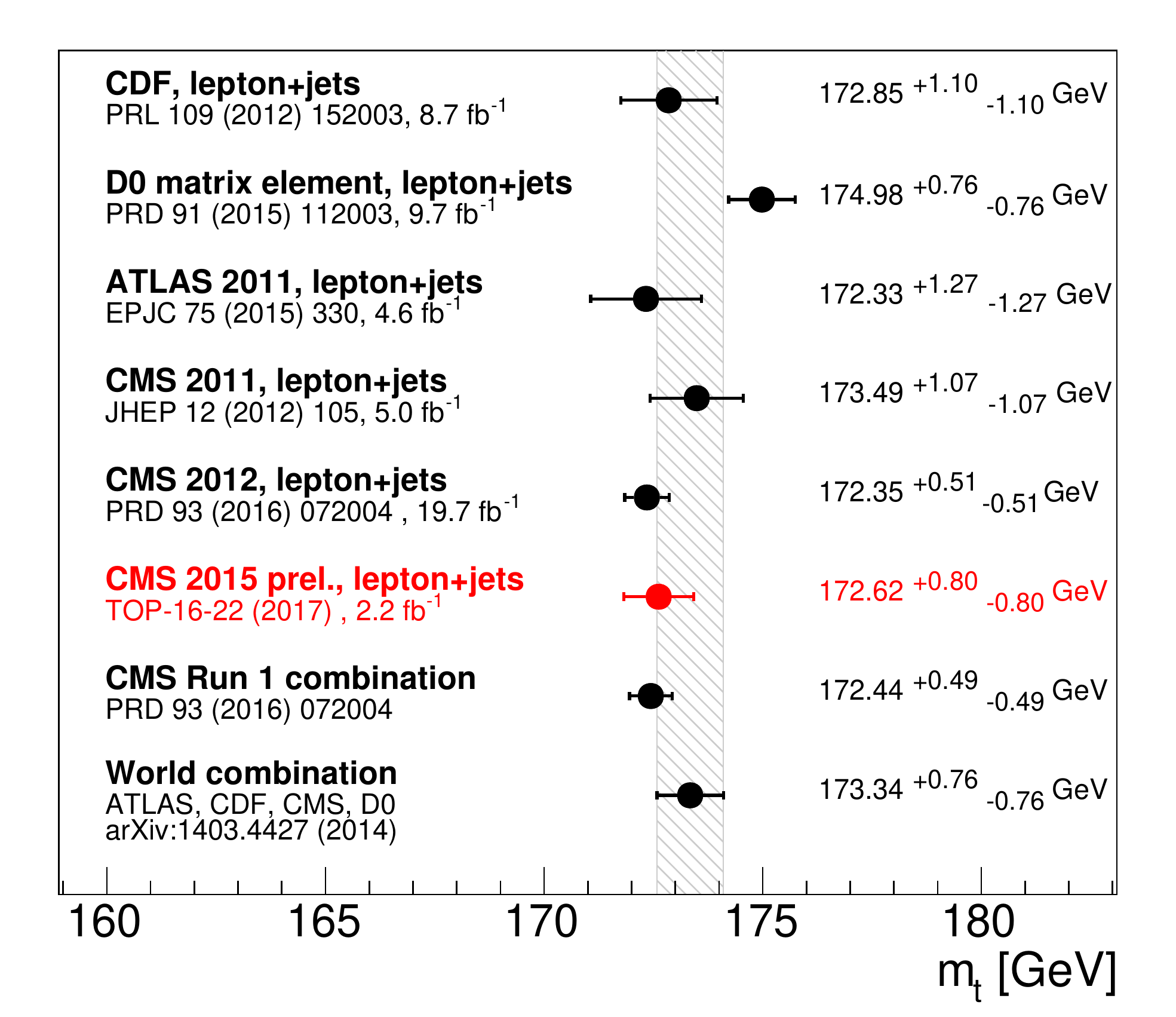}}
\vspace*{8pt}
\caption{Comparison of the first 13~TeV top quark mass measurement (in red) with previous measurements with lepton+jets final states and the CMS and world top quark mass averages~\cite{massR2}.\protect\label{tmassp}}
\end{figure}

In order to properly measure top quark properties, large sets of well-understood data are needed, and for that reason the time for such measurements based on Run~2 data has yet to arrive. Two fundamental measurements, those of the mass and the width of the top, however, are already available in Run~2. The measurement of the top quark width~\cite{wd} sets a bound of $0.6\leq \Gamma_t \leq 2.5$~GeV (the NLO SM prediction is 1.35~GeV) at 13~TeV comparing data with different $\Gamma_{t}$ hypothesis. The first mass measurement~\cite{massR2} at 13~TeV, in the lepton+jets channel, used a classic template method and obtained a value of $\mathrm{m_t}=172.62\pm0.38\rm{(stat.+JSF)}\pm0.70\rm{(syst.)}$~GeV. This result is consistent with the Run~1 measurements, as it can be seen in Fig.~\ref{tmassp}.

Rare top quark production mechanisms will be studied extensively during Run~2 and beyond. Associated production of $t\bar{t}V$ is particularly interesting, as it is sensitive to BSM effects, anomalous couplings, and leading effective field theory (EFT) operators. At 13~TeV, $t\bar{t}V$ was studied~\cite{ttVR2} in three channels defined by the leptons present in the final state. The analysis was based on a simultaneous fit across several regions of the phase space using multivariate discriminants. With the full 2016 dataset, $t\bar{t}Z$ is observed with 9.9$\sigma$ of significance and $t\bar{t}W$ with 5.5$\sigma$, establishing the two processes independently for the first time. The first studies of four top-quark production have also been performed at 13~TeV, and the latest supersymmetry (SUSY) analysis in final states with same-sign dileptons~\cite{susyss}, using 35.9$fb^{-1}$ of data sets an observed limit 4.6 times the SM for $\sigma_{t\bar{t}t\bar{t}}$. 

Top quark physics at the LHC has entered the high precision regime and the modelling, level of accuracy of the theory predictions, and proper evaluation of systematic uncertainties are taking a central stage. So far, there are no clear tensions with the SM predictions in the top quark sector at the LHC experiments, with one exception. In differential measurements there is generally good agreement with NNLO predictions and NLO generators. However, the top quark $p_T$ spectrum is persistently found to be softer in data than in the simulation. This effect was observed during Run~1, and it is still present in Run~2. This effect is observed in all channels, in $t\bar{t}$ and single top t-channel production, and beyond the TeV scale in the boosted regime. The top quark $p_T$ description is improved by higher order calculations (NNLO providing a better description than NLO) and simulation at higher orders (NLO generators seeing a smaller discrepancy than LO). More data and additional studies, such as double differential distributions, are expected to shed light on this topic in the near future. 

\section{Top Quark and Higgs Boson physics: the Hunt is on!}

The top quark and the Higgs boson have a special relationship. The Higgs boson couples to mass, and therefore its coupling to the top quark, whose mass is close to the EWK symmetry breaking scale, is expected to be very strong in the SM. Moreover, the precise measurement of the top quark mass and the W boson mass set constraints on the expected mass of a SM Higgs boson via EWK fits~\cite{gfitter}. Additionally, the top quark radiative corrections can drive the Higgs boson self-coupling towards negative values, potentially leading to an unstable vacuum that would need new phenomena to be stabilized. Hence, a precise measurement of the mass of the Higgs boson and the mass of the top quark would allow for conclusions on the fate of the Universe via the analysis of the vacuum stability~\cite{vacuum}.

The top quark has an impact in the Higgs sector: it is involved in the main Higgs boson production mode at the LHC, gluon-gluon fusion, and in one of the main Higgs boson decays, Higgs to $\gamma\gamma$, in both cases via loop at leading order. This makes possible to experimentally constrain the top-Higgs coupling, an excellent probe of the SM consistency, by studying the gluon-gluon fusion production and H$\rightarrow \gamma\gamma$ decay. Within uncertainties, the current measurement of this coupling is so far consistent with the SM expectations. However, a direct measurement of this coupling can only be performed in processes where top quark and Higgs boson are produced together. The coupling cannot be assessed by measuring Higgs decays to top quark pairs because the top quark is heavier than the Higgs boson.

\begin{figure}[h]
\centerline{\includegraphics[width=2.0in]{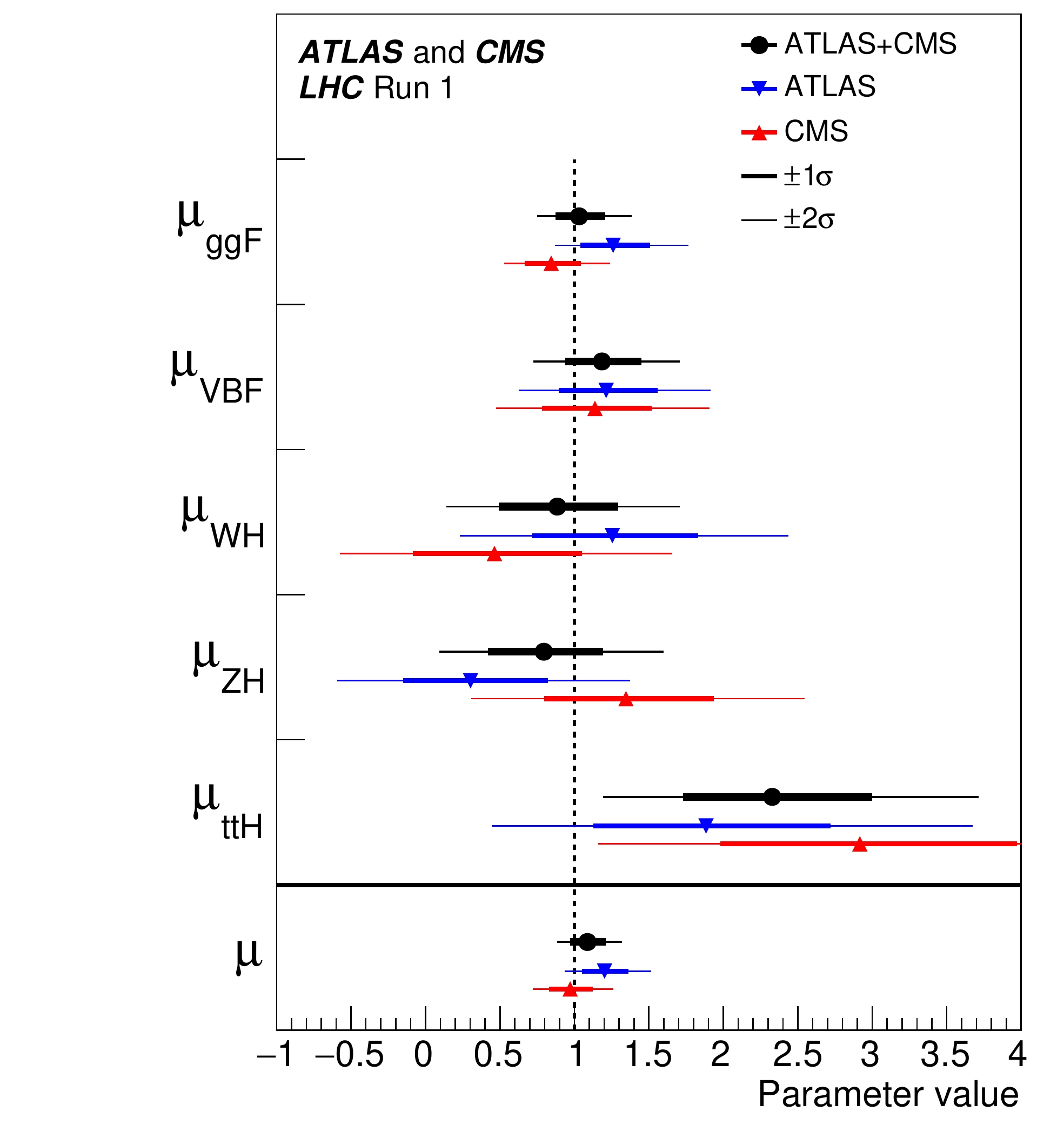}
\includegraphics[width=2.0in]{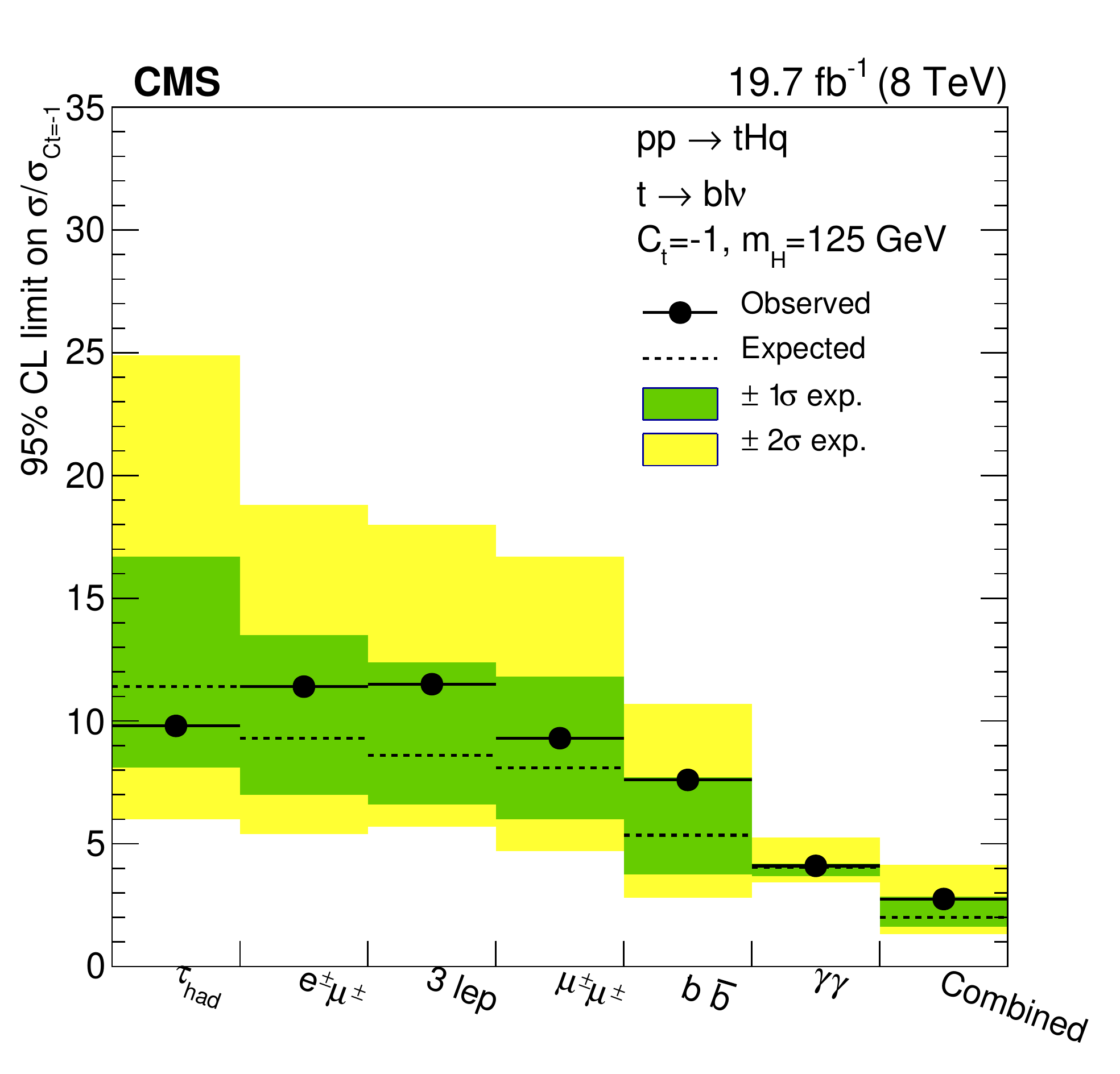}}
\vspace*{8pt}
\caption{Left: Best fit results for the production signal strengths for the combination of ATLAS and CMS data including $t\bar{t}$H~\cite{lhccoup}. Right: The 95\% CL upper limits on the excess event yields predicted by the enhanced tHq cross section and Higgs boson to diphoton branching fraction for $\kappa_t=-1$. The limits are normalized to the $\kappa_t=-1$ prediction, and are shown for each analysis channel, and combined~\cite{thrun1}. \protect\label{r1tth}}
\end{figure}

The straightforward channel to directly study the top-Higgs coupling is therefore Higgs produced in association with a top quark pair, $t\bar{t}$H. With Run~1 data, this production was studied in all possible decay channels~\cite{tthrun1}: H$\rightarrow b\bar{b}$, H$\rightarrow \gamma\gamma$, H$\rightarrow$WW, H$\rightarrow$ZZ, and H$\rightarrow \tau\tau$. All channels studied are statistically limited -since $t\bar{t}$H is the sixth process in terms of production cross section at the LHC- but the results are still important and can be combined. The best-fit value for the signal strength of $t\bar{t}$H production in Run~1, obtained in combination of the ATLAS and CMS results~\cite{tthlhc}, was found to be 2.8, driven by an excess above the background-only hypothesis of 3.4 standard deviations (2$\sigma$ upward deviation with respect to the SM prediction). The measurements of the Higgs signal strength separated by channel, including $t\bar{t}$H are presented in Fig.~\ref{r1tth} (Left).

There is another process in which top and Higgs are produced together that is sensitive to the relative sign of the top-Higgs coupling, something that $t\bar{t}H$ is blind to: single top production in association with a Higgs boson (tH: tHq and tHW). The tH production is highly suppressed in the SM but its rate could be affected by anomalous top-Higgs couplings. In particular, a negative value of the Higgs boson coupling to the top quark (assuming the SM value for the coupling to vector bosons) could increase the tHq cross section by a factor of 15. Single top plus Higgs is then sensitive to the size of the coupling, and to the relative sign between $\kappa_V$ and $\kappa_t$. Negative values $\kappa_t$ are excluded by the current measurements in gluon-gluon fusion and H$\rightarrow \gamma\gamma$, but this exclusion is made under the assumption that no new particles contribute to the loops, and that there are no BSM decays.  

The production of tHq has very characteristic signatures: a Higgs boson decay, a top quark decay, and a forward light jet, which is the classic single top t-channel handle. The CMS Collaboration studied tHq with Run~1 data~\cite{thrun1} in as many channels as $t\bar{t}$H: H$\rightarrow b\bar{b}$, H$\rightarrow \gamma\gamma$, H$\rightarrow$WW, H$\rightarrow$ZZ, and H$\rightarrow \tau\tau$. The SM signal is too small to be measured, but 95\% CL upper limits were set on anomalous tHq production ($\kappa_t=-1.0$, $\kappa_V=1.0$) as it is presented in Fig.~\ref{r1tth} (Right). 

At 13~TeV, the production cross sections of both $t\bar{t}$H and tH, grow substantially ($t\bar{t}$H goes from 0.13~pb at 8~TeV to  0.51~pb at 13~TeV), more than other Higgs production modes, and more than the main backgrounds that can produce similar final states. The LHC Run~2 is optimal for the study of these processes, and will allow for a precise determination of both the magnitude and the sign of the top quark Yukawa coupling. 

Preliminary $t\bar{t}$H and tH results in different decay modes by the CMS experiment performed on the current 13~TeV dataset are already available and will be briefly presented below. 

The $t\bar{t}H$, $H\rightarrow b\bar{b}$~\cite{tthbb} channel has been studied in Run~2 using a set of event categories based on the number of (b) jets in the final state. Sophisticated multivariate techniques (BDT, Matrix Element Methods) are used in the signal enriched regions. The top quark production process $t\bar{t}$+$b\bar{b}$ is a very important background to control in the analysis.

The high-resolution channels, $H\rightarrow \gamma\gamma$~\cite{tthgg} and $H\rightarrow ZZ\rightarrow 4l$~\cite{tthzz}, suffer from a small branching fraction that, combined with the low production rate of $t\bar{t}$H, results in very low signal yields. Both channels have been explored with 13~TeV data with limited sensitivity.

Final states with same-sign dileptons, trileptons or four leptons~\cite{tthlep} are sensitive to $t\bar{t}$H, H$\rightarrow WW$ and to a lesser extent to H$\rightarrow ZZ$ and H$\rightarrow \tau\tau$. In this kind of `multilepton' analysis, to control the $t\bar{t}V$ background is crucial. The analysis uses event categories based on the number of leptons, hadronically decaying taus, lepton flavor, charge, number of (b) jets; and, in the same way as in H$\rightarrow b\bar{b}$, uses multivariate techniques. The $t\bar{t}H$ multilepton is at the moment the most sensitive decay for $t\bar{t}H$: with 35.9$\rm{fb^{-1}}$ CMS achieves evidence  for $t\bar{t}$H production with a significance of 3.3$\sigma$ (2.5 expected).  

\begin{figure}[h]
\centerline{\includegraphics[width=2.0in]{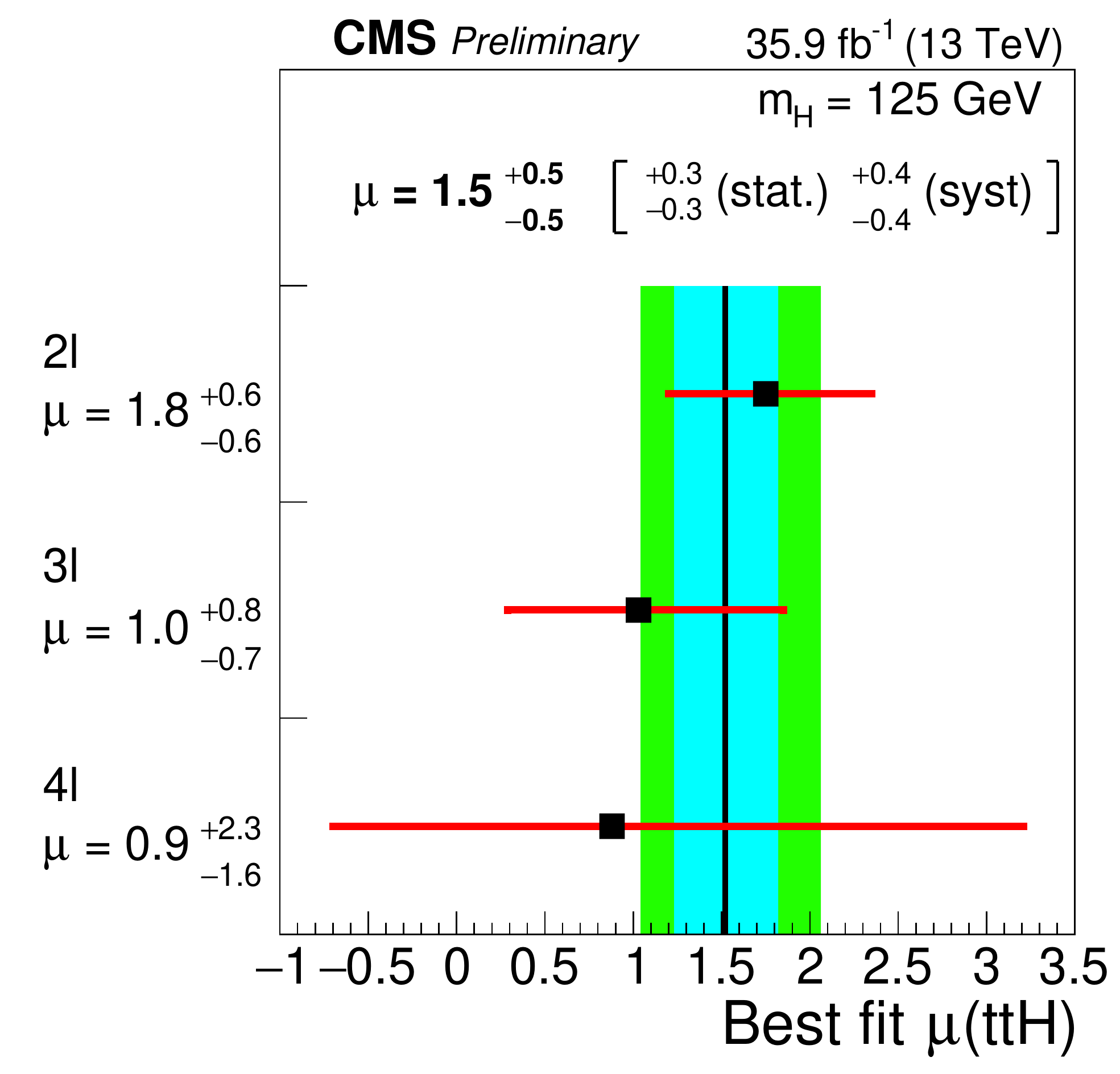}
\includegraphics[width=2.0in]{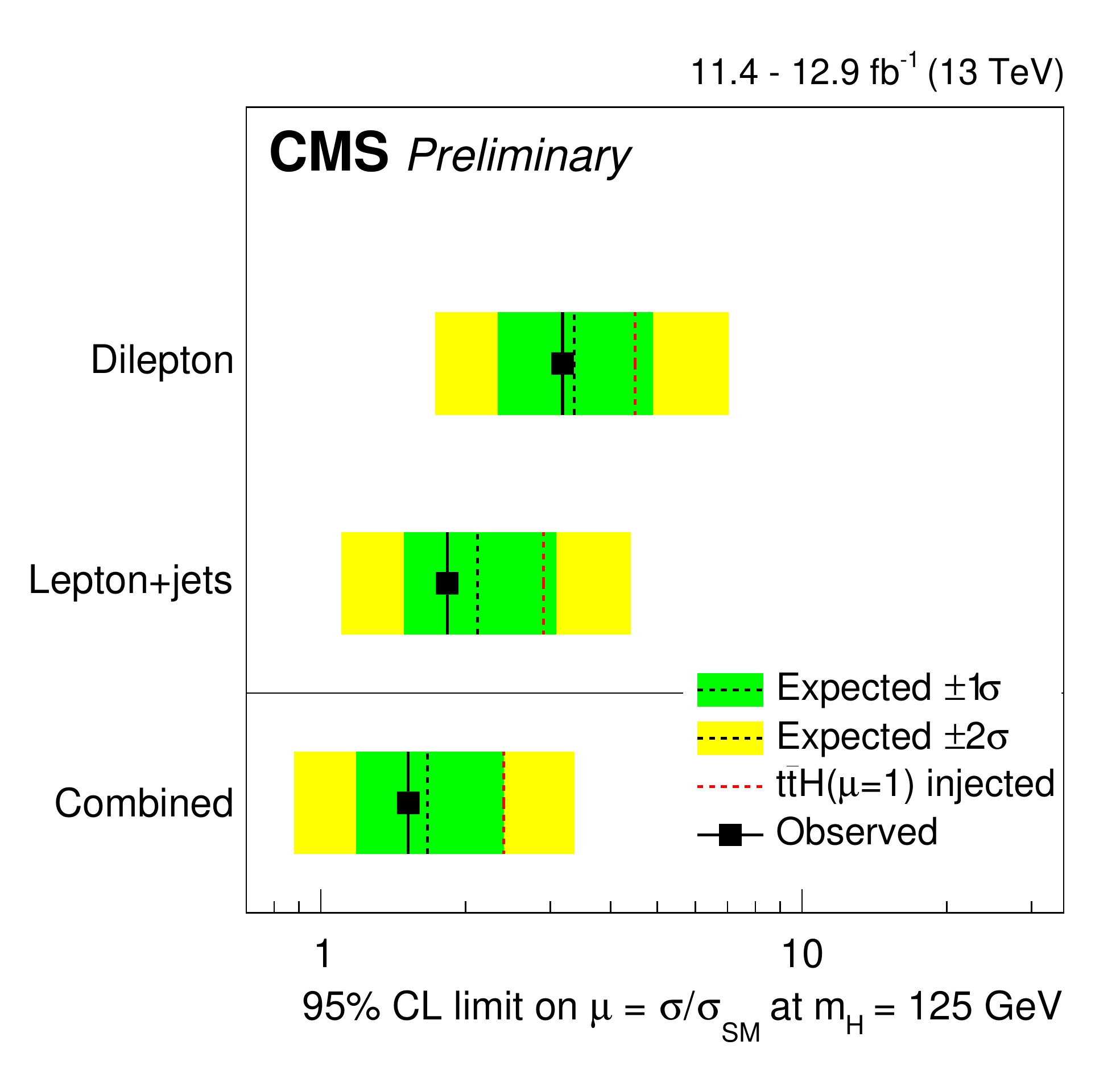}}
\vspace*{8pt}
\caption{Left:  Best fit signal strength for the 2016 $t\bar{t}$H multilepton analysis in different channels~\cite{tthlep}. Right: Median expected and observed 95\% CL upper limits on $\mu_{ttH}$ in the H$\rightarrow b\bar{b}$ analysis~\cite{tthbb}. The expected limits are displayed together with $\pm1\sigma$ and $\pm2\sigma$ confidence intervals. Also shown are the limits in case of an injected signal of $\mu=1$. \protect\label{mutth}}
\end{figure}

Halfway through Run~2 the CMS Collaboration has produced an extensive collection of results. As presented in Fig.~\ref{mutth} the most sensitive channel still measures a slightly larger signal strength than the expected, but more data is needed to understand if this effect is physical.

\begin{figure}[h]
\centerline{\includegraphics[width=2.0in]{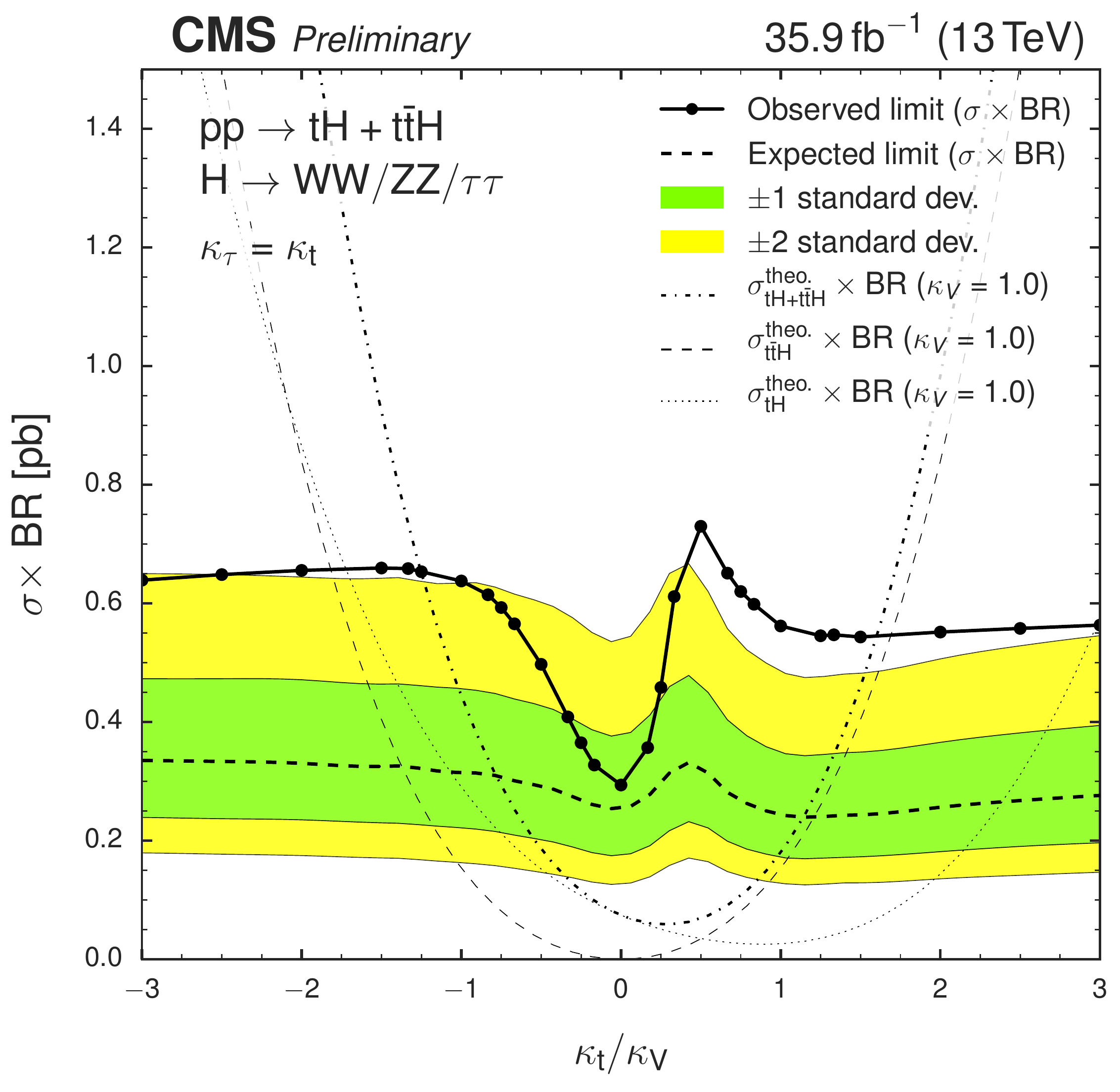}
\includegraphics[width=2.0in]{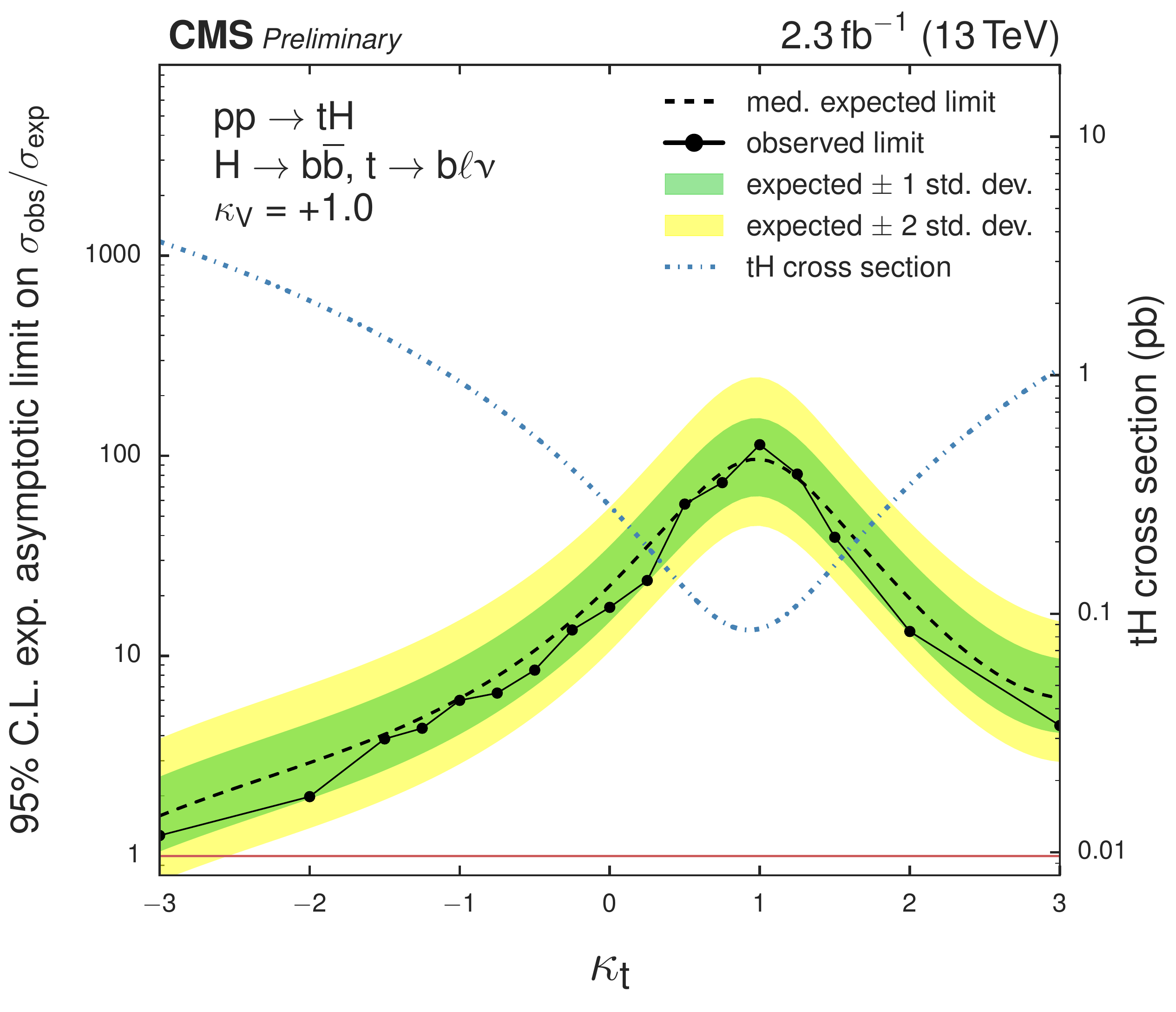}}
\vspace*{8pt}
\caption{Left: Observed and expected 95\% CL upper limit on the tH+$t\bar{t}$H cross section times branching fraction into WW, $\tau\tau$, and ZZ for different values of the coupling ratio $\kappa_t$/$\kappa_V$~\cite{thl}.
Right: Upper limits on tH assuming a SM coupling to vector bosons in the $b\bar{b}$ decay~\cite{thbb}. 
\protect\label{thp}}
\end{figure}

Preliminary results covering single top plus Higgs with Run~2 data are also available, with a similar approach to $t\bar{t}$H in multilepton~\cite{thl} (dilepton with same-sign, and trileptons) and $b\bar{b}$ decays~\cite{thbb}. These searches explicitly target different $\kappa_t$-$\kappa_V$ coupling scenarios, keeping as benchmark $\kappa_t=-1.0$ (with $\kappa_V=1.0$). In the multilepton analysis, performed with the full 2016 dataset, values of $\kappa_t$ outside the range of -1.25 to +1.60 are excluded at 95\% CL assuming $\kappa_V=1.0$, as can be seen in Fig.~\ref{thp} (Left).

The number of top quark and Higgs boson results in CMS is increasing with time and center of mass energy. With an estimate of 100$fb^{-1}$ of proton-proton collisions at 13~TeV to look forward to at the end of Run~2 it is very likely that $t\bar{t}$H will be observed and the top-Higgs coupling measured in the very near future.  

A large rate of top quark plus Higgs boson (single or otherwise) events could also be a sign for other interesting BSM phenomena, such as new physics showing up as Higgs boson mediated flavor-changing neutral-currents (FCNH), or direct production of heavy new particles -as predicted in composite and little Higgs models (e.g. single or pair-produced vector-like T quarks decaying to top quarks and Higgs bosons, T$\rightarrow$tH). 

There are in fact many more exotic processes that lead to final states with top quarks and Higgs bosons which have been studied during Run~1. Extensions of the SM such as two-Higgs-doublet model R-parity violating minimal supersymmetric standard model (MSSM) models, and warped extra dimensions predict a larger branching fraction of FCNH  t$\rightarrow$Hq (smaller than $10^{-13}$ in the SM). The Higgs sector of the MSSM considers two charged Higgs bosons $\mathrm{H^{\pm}}$. For a low mass charged Higgs, dominant decays also share top-Higgs signatures: t$\rightarrow H^{+}b$, for $tan\beta < 1$ in some scenarios, $H^{+}\rightarrow cs$. No sign of FCNH or charged Higgs bosons have been found so far at the LHC. 

Processes with top quarks and Higgs bosons that are scrutinized in search for new phenomena are in general: rare decays (e.g. top decaying to Higgs in FCNH processes); heavy resonances (e.g. heavy Higgs partners decaying to $t\bar{t}$, or resonances decaying to top and Higgs); new particles produced with top and/or Higgs; new particles decaying to top and/or Higgs; top partners decaying to top quarks and/or Higgs bosons; and Dark matter produced together with or coupling to top quarks and/or Higgs bosons. 

With the new data, the top quark and Higgs boson sectors will continue providing high precision measurements to push the SM limits. In the absence of strong indicators of BSM phenomena, such as new particles, the properties of the Higgs boson and the top quark could still hold the answers to the open questions of the SM. New physics at 1 TeV could for example show up as deviations in the Higgs boson couplings of at most 10\%. Sensitivity to compositeness or SUSY effects could appear below 5\%, making a clear case for ultimate precision.

\section{Summary} 

With unprecedented collision energies and integrated luminosity, the LHC experiments are in a privileged position to study the Higgs boson and the top quark. During the Run~1 of the LHC, at 7 and 8~TeV, the Higgs boson was not only discovered but characterized in detail. The top quark legacy from the same period puts the LHC experiments at the level of the Tevatron, where the top quark was observed for the first time, achieving outstanding accuracy in the measurement of central SM parameters, such as the top mass, determined with a precision of 0.3\%.  

No other new particles have been found at the LHC, but the fundamental open questions of the SM still require new physics. Both the top quark and the Higgs boson are good candidates to participate in new physics scenarios, and considering them together could be a powerful tool to explore the unknown. Among the most relevant SM quantities to measure at the LHC, the Higgs coupling to the top quark is particularly interesting. 

The conditions of the LHC Run~2 are especially suited to study processes in which top quarks and Higgs bosons are produced together and already many preliminary results are available. The measurement of the coupling of the Higgs boson to the top quark is within reach.  Analyses limited by statistics during Run~1 will not be so with Run~2 data. While so far the LHC measurements are compatible with the predictions within uncertainties, new phenomena could be hiding in deviations as small as just a few percent, motivating the LHC to reach for the best precision possible.   

Only time will tell if the exploration of the top quark and the Higgs boson will hold the answers to some of the fundamental questions we are trying to address, or if it will open new, unexpected, ones. Either way, such studies, with the data the LHC will keep providing in the next years, hold great promise.

\appendix

\section*{Acknowledgments}

A big thank you to Ken Bloom and Ilya Kravchenko for the useful comments and suggestions. I wish to thank also Pedro Silva, Maria Aldaya, and Elias Coniavitis for reading this paper. Finally, I would like to express my gratitude to Boaz Klima and the Fermilab CMS group that invited me to give this seminar.

This material is based upon work supported by the National Science Foundation under Grant No. 1607202.


\begin{thebibliography}{0}

\bibitem{Higgs1964} P.~W.~Higgs, {\it Phys.\ Lett.} {\bf 12}, 132 (1964).

\bibitem{top1} F.~Abe {\it et al.} (CDF Collaboration) {\it Phys.\ Rev.\ Lett.} {\bf 74}, 2626 (1995).

\bibitem{top2} S.~Abachi {\it et al.} (D0 Collaboration) {\it Phys.\ Rev.\ Lett.} {\bf 74}, 2632 (1995).

\bibitem{higgs1} G.~Aad {\it et al.} (ATLAS Collaboration) {\it Phys.\ Lett.\ B} (2012), Vol.~716, pp.~1--29

\bibitem{higgs2} S.~Chatrchyan {\it et al.} (CMS Collaboration) {\it Phys.\ Lett.\ B} (2012), Vol.~716, pp.~30--61

\bibitem{lhcb} (LHCb Collaboration) {\it LHCB-PAPER-2017-013} (2017).

\bibitem{Perl} M.~L.~Perl, {\it Phys. perspect.} (2004), Vol.~6, pp.~401--427.

\bibitem{hrun1} V.~Khachatryan {\it et al.} (CMS Collaboration) {\it Eur.\ Phys.\ J.\ C} {\bf 75}, no. 5, 212 (2015).

\bibitem{spin} V.~Khachatryan {\it et al.} (CMS Collaboration), {\it Phys.\ Rev.\ D} {\bf 92}, no. 1, 012004 (2015).

\bibitem{spinrun2} (CMS Collaboration), {\it Preliminary}, PAS-HIG-17-011 (2017).

\bibitem{cmsmass}  V.~Khachatryan {\it et al.}  (CMS Collaboration), {\it Eur.\ Phys.\ J.\ C} {\bf 75}, no. 5, 212 (2015).

\bibitem{lhccoup} G.~Aad {\it et al.} (ATLAS and CMS Collaborations), {\it JHEP} {\bf 08}, 045 (2016).

\bibitem{lhcmass}  G.~Aad {\it et al.} (ATLAS and CMS Collaborations), {\it Phys.\ Rev.\ Lett.} {\bf 114}, 191803 (2015).

\bibitem{hwidth} V.~Khachatryan {\it et al.} (CMS Collaboration), {\it Phys.\ Rev.\ D} {\bf 92}, no. 7, 072010 (2015).

\bibitem{HIG-16-041} (CMS Collaboration), {\it Preliminary}, PAS-HIG-16-041 (2017).

\bibitem{xsec}   V.~Khachatryan {\it et al.} (CMS Collaboration), {\it JHEP} {\bf 08}, 029 (2016).

\bibitem{diff}  V.~Khachatryan {\it et al.} (CMS Collaboration), {\it Eur.\ Phys.\ J.\ C} {\bf 77}, no. 1, 15 (2017).

\bibitem{tW}  S.~Chatrchyan {\it et al.} (CMS Collaboration), {\it Phys.\ Rev.\ Lett.} {\bf 112}, no. 23, 231802 (2014).

\bibitem{ttV}  V.~Khachatryan {\it et al.} (CMS Collaboration), {\it JHEP} {\bf 01}, 096 (2016).

\bibitem{topmass}  V.~Khachatryan {\it et al.} (CMS Collaboration), {\it Phys.\ Rev.\ D} {\bf 93}, no.7,  072004 (2016).

\bibitem{xsecEA}  V.~Khachatryan {\it et al.} (CMS Collaboration), {\it Eur.\ Phys.\ J.\ C} {\bf 77}, 172 (2017).

\bibitem{diffR2}  V.~Khachatryan {\it et al.} (CMS Collaboration), {\it Phys.\ Rev.\ D} {\bf 95}, no. 9, 092001 (2017).

\bibitem{tchEA}  A.~M.~Sirunyan {\it et al.} (CMS Collaboration), {\it Accepted for publication by Phys.\ Lett.\ B}.

\bibitem{wd} (CMS Collaboration), {\it Preliminary}, PAS-TOP-16-019 (2016).

\bibitem{massR2} (CMS Collaboration), {\it Preliminary}, PAS-TOP-16-022 (2017).

\bibitem{ttVR2} (CMS Collaboration), {\it Preliminary}, PAS-TOP-17-005 (2017).

\bibitem{susyss}  A.~M.~Sirunyan {\it et al.} (CMS Collaboration), {\it  Submitted to Eur. Phys. J. C.}

\bibitem{gfitter} M.~Baak {\it et al.} (Gfitter Group) {\it Eur.\ Phys.\ J.\ C} {\bf 74}, 3046 (2014).

\bibitem{vacuum}  A.~V.~Bednyakov, B.~A.~Kniehl, A.~F.~Pikelner and O.~L.~Veretin, {\it Phys.\ Rev.\ Lett.}  {\bf 115}, no. 20, 201802 (2015).

\bibitem{tthrun1} V.~Khachatryan {\it et al.} (CMS Collaboration), {\it JHEP} {\bf 10}, 106 (2014).

\bibitem{tthlhc} G.~Aad {\it et al.} (ATLAS and CMS Collaborations), {\it JHEP} {\bf 08}, 045 (2016).

\bibitem{thrun1} V.~Khachatryan {\it et al.} (CMS Collaboration), {\it JHEP} {\bf 06}, 177 (2016).

\bibitem{tthbb} (CMS Collaboration), {\it Preliminary}, PAS-HIG-16-038 (2016).

\bibitem{tthgg} (CMS Collaboration), {\it Preliminary}, PAS-HIG-16-020 (2016).

\bibitem{tthzz} (CMS Collaboration), {\it Preliminary}, PAS-HIG-16-041 (2016).

\bibitem{tthlep} (CMS Collaboration), {\it Preliminary}, PAS-HIG-17-004 (2017).

\bibitem{thl} (CMS Collaboration), {\it Preliminary}, PAS-HIG-17-005 (2017).

\bibitem{thbb} (CMS Collaboration), {\it Preliminary}, PAS-HIG-16-019 (2016).


\end{thebibliography}
\end{document}